\documentclass[final,english]{bullsrsl}[2022/06/15]

\usepackage[latin1]{inputenc}
\usepackage[T1]{fontenc}

\usepackage{natbib} 
\usepackage{graphicx}

\begin{document}
\title{Study of instabilities and outbursts in luminous blue variables \\ \textit{AF And} $\&$ R\,127}

\author[affil={1}, corresponding]{Abhay Pratap}{Yadav}
\author[affil={1}]{Sugyan}{Parida}
\author[affil={2}]{Yogesh Chandra}{Joshi}
\author[affil={2}]{Santosh}{Joshi}
%\author[affil={4}]{Lotta}{Lothardis}
\affiliation[1]{Department of Physics and Astronomy, National Institute of Technology, Rourkela - 769008, India }
\affiliation[2]{Aryabhatta Research Institute of Observational Sciences, Manora Peak, Nainital - 263002, India}
%\affiliation[3]{Institute of Improbability, Impossible City}
%\affiliation[4]{Double Blind Testing Inc., Great Opportunities}
\correspondance{yadavap@nitrkl.ac.in}
\date{10th June 2023}
\maketitle

% \author[affil1]{FirstName (+ MiddleInitials if necessary)}{FamilyName}
% \author[affil2]{...}{}
% \equalcontribauthor[]{}{} % Maximum two --> counter
% \consortium[affil]{Consortium Name}
% With consortium: affiliation will be set to "See Appendix 1 for a full
% list of consortium members and their respective affiliations
% \affiliation[affil1]{...}
% \affiliationq[affil2]{...}

% \correspondence[]{}
% No explicit corresponding author: use first author
% 

% Abstract of the paper in the same language as the paper
\begin{abstract}

Luminous blue variables (LBVs) are
evolved massive stars close to the Eddington limit, with a distinct
spectroscopic and photometric variability having unsteady mass-loss rates.
These stars show a considerable change in their surface temperature from
quiescent to outbursts phase. The cause of irregular variability and unsteady
mass-loss rate is not properly understood.
Here we present the result of linear stability analysis in two LBVs \textit{AF And}
and R\,127 during their quiescent and outburst phase. We note that several
modes are unstable in the models of the considered LBVs. Mode interaction is frequent 
in the modal diagrams for the models of both LBVs.
For \textit{AF And}, number of instabilities increase in models having temperature below 
15000 K.  
The found
instabilities may be linked with the observed irregular variabilities and 
surface eruptions. Observational facilities of Belgo-Indian Network for Astronomy and Astrophysics (BINA) will be very
beneficial to study the spectroscopic and photometric behavior of the
considered LBVs.
%The whole document is formatted with the Times New Roman 
%font and a line spacing of 1.15 (double line spacing with the ``manuscript'' option).
%The font size is 12 everywhere except in some titles.
%Text paragraphs are justified to the left and right margins.
%The \textsf{bullsrsl} \LaTeX\ class takes care of all these 
%requirements; it is sufficient to use the adequate sectioning commands.
\end{abstract}


%\section{Section -- Level 1 title (Times New Roman, bold, 14 pts)}
\section{Introduction}
Luminous blue variables (LBVs) are the late evolutionary transient phase of most massive 
stars which shows a very strong photometric and spectral variability at time scales 
ranging from months to years \citep{humphreys_1994, van_2001}.  Generally stars having
mass equal or greater than 21 M$_{\odot}$ pass through this temporary but quite 
influential phase of evolution \citep{weiss_2020}.  During this phase, stars lose 
a considerable amount of mass via surface eruptions and other unsteady mass-loss processes.
LBVs are believed to be the progenitor of Wolf- Rayet stars \citep[e.g.,][]{groh_2014}. However, recent studies are 
also indicating that some LBVs may be the progenitors of peculiar
supernovae \citep{vink_2006,groh_2013, smith_2017}. Irregular variabilities and surface
eruptions are the main properties of LBVs but the origin of variabilities and eruptions 
are not properly understood. In this context, the present work is a preliminary step to 
study instabilities in models of two very famous LBVs \textit{AF And} and R\,127. 

\textit{AF And} is one of the most luminous stars in Andromeda. During 1917 to 1953 at least five
major eruptions were observed. \citet{yogesh_2019} have pointed out the surface temperature 
of this star changes in the range of 30000 K to 7000 K. During the quiescent phase AF 
And has mass-loss rate of 2.2 x 10$^{-4}$ M$_{\odot}$ yr$^{-1}$ and wind terminal 
velocity is found to be in the range of 280 to 300 km s$^{-1}$.
Radcliffe (R) 127 is another LBV located in the evolved cluster NGC 2055 of Large Magellanic
Cloud \citep{walborn_1991, heydari_2003}. While passing through the quiescent to outburst
phase surface temperature of R\,127 changes from approximately 30000 K to 9000 K. 

Method and parameters to construct the models of \textit{AF And} $\&$ R\,127 are mentioned in section \ref{models}.
Linear stability analysis and obtained results are presented in section \ref{lsa_r}, followed by 
discussion and conclusion in section \ref{d_and_c}. 
%\subsection{Subsection -- Level 2 title (Times New Roman, bold, 12 pts)}
\section{Models of \textit{AF And} and R\,127} \label{models}

To construct models of the considered stars, parameters such as mass, luminosity, range of surface temperature 
and chemical compositions are required. In case of several luminous blue variables, due to 
uncertainty in distance measurement, luminosity is not accurately known. Similarly, mass of several LBVs is 
uncertain as many of them are single field stars. In absence of precise value of mass and luminosity, it is quite 
challenging to model LBVs and other very massive stars. To tackle this issue, instead of considering a 
single model, a sequence of models are generally used \citep{yadav_2016, yadav_2021}. 
To consider the quiescent and outburst phase, we have adopted a sequence of models 
for both \textit{AF And} as well as R\,127 having surface temperature in the range of 32000 K to 9000 K.       

For model construction, we have integrated the stellar structure equations as an initial value problem 
from the surface upto interior having temperature of 10$^{7}$ K. Stefan-Boltzmann's law and photospheric pressure are 
used as boundary conditions to perform the inwards integration upto temperature T = 10$^{7}$ K. Rotation and magnetic fields 
are disregarded to simplify the modelling problem. Schwarzschild's criteria is used for the onset of convection. 
For all the considered models, OPAL opacity tables \citep{rogers_1992,rogers_1996} are used. For the convection,
mixing length theory \citep{bohm_1958} with mixing length parameter $\alpha$ = 1.5 is used.         

\begin{figure}
\centering $
%\large
\begin{array}{c}
 \scalebox{0.8}{ \input{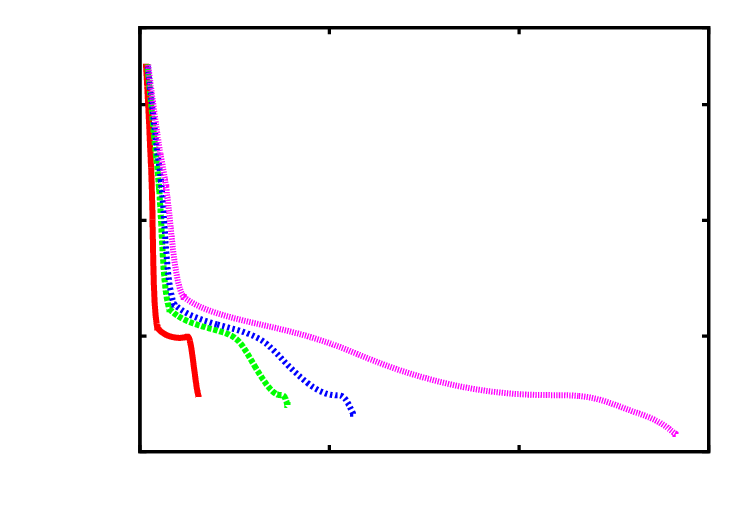} } \\
 \end{array}$
 \caption{Density profile of the models of \textit{AF And} with mass of 75
M$_{\odot}$ and log L/L$_{\odot}$ = 6.25. Density inversion can be noticed in the
small region of the envelopes.}
 \normalsize
 \label{dens}
 \end{figure}

Models of \textit{AF And} $\&$ R\,127 have been constructed with solar chemical composition ( X = 0.70, Y = 0.28 and Z = 0.02 where X, Y and Z are the mass fractions of hydrogen, helium and heavier elements, respectively ) using mass, luminosity and the 
range of surface temperature mentioned in the Table.\ref{tab:my-table}. Density as a function of radius for four different models of \textit{AF And} is given in Fig.\ref{dens}. Models having lower surface 
 temperature have quite extended envelope compared to models having surface temperature $\ge$ 25000 K. Presence of 
 density inversion can also be noticed in Fig.\ref{dens}.  
\begin{center}
 
\begin{table}[h] 
\centering
\begin{tabular}{|c|c|c|c|}
\hline
Star Name & \begin{tabular}[c]{@{}c@{}}Mass \\ {[}M$_{\odot}${]}\end{tabular} & \begin{tabular}[c]{@{}c@{}}Luminosity\\ {[}log L/L$_{\odot}${]}\end{tabular} & \begin{tabular}[c]{@{}c@{}}Surface Temperature Range\\ {[}K{]}\end{tabular} \\ \hline
\textit{AF And}    & 75                                                                   & 6.25                                                                         & 32000 K to 9000K                                                            \\ \hline
R\,127     & 55                                                                   & 6.15                                                                         & 32000 K to 9000K                                                            \\ \hline
\end{tabular}
\caption{Mass, luminosity and surface temperature range for the considered LBVs. The value of luminosity
is adopted from \citet{jiang_2018}.}
\label{tab:my-table}
\end{table}
\end{center}

%All paragraphs in the main text are indented by about 0.75\,cm.
% The spacing between paragraphs is 6\,pts (this is set in the class file -- no need to worry).
%The figures and tables should be placed close to the paragraph where they
%are referred to for the first time.
%They might be moved by the publisher during the final formatting.

%Footnotes or endnotes are \emph{not allowed}. 

%\subsubsection{Subsubsection -- Level 3 title (Times New Roman, italics, 12 pts)}
\section{Linear Stability Analysis and Results} \label{lsa_r}
For the case of radial perturbation, linearized pulsation equations form a fourth order boundary eigenvalue problem. 
In the present study, we have adopted the standard form of pulsation equations as mentioned 
by \citet{baker_1962} and \citet{gautschy_1990a}.For a detailed discussion on the used pulsation equations and boundary conditions, the reader
is referred to \citet{gautschy_1990a}. To solve this boundary eigenvalue problem, 
the Riccati method is used. The 
obtained eigenfrequencies are complex ($\sigma$ = $\sigma_r$ + $i$ $\sigma_i$) where 
real part ($\sigma_r$) is linked
with the pulsation period and imaginary part ($\sigma_i$) is indicating excitation or 
damping. Modes with $\sigma_i$ < 0 
are excited while the modes having $\sigma_i$ > 0 are damped. In the present study, eigenfrequencies are 
normalized with the global free fall timescale ($\sqrt{R^3/3\,G\,M}$; where $R$, $G$ and $M$ are the stellar radius, gravitational constant and stellar mass, respectively) in such a way that the eigenfrequencies become dimensionless.

\begin{figure*}
\centering $
\large
\begin{array}{cc}

   \scalebox{0.65}{ \input{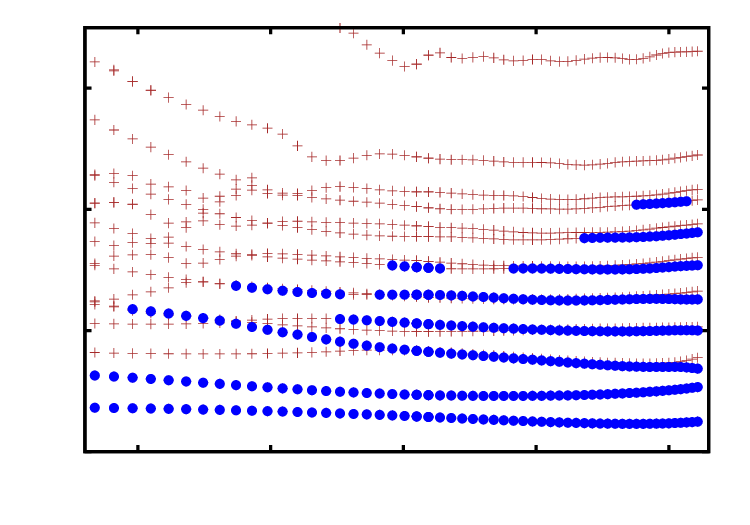} } 
   \scalebox{0.65}{ \input{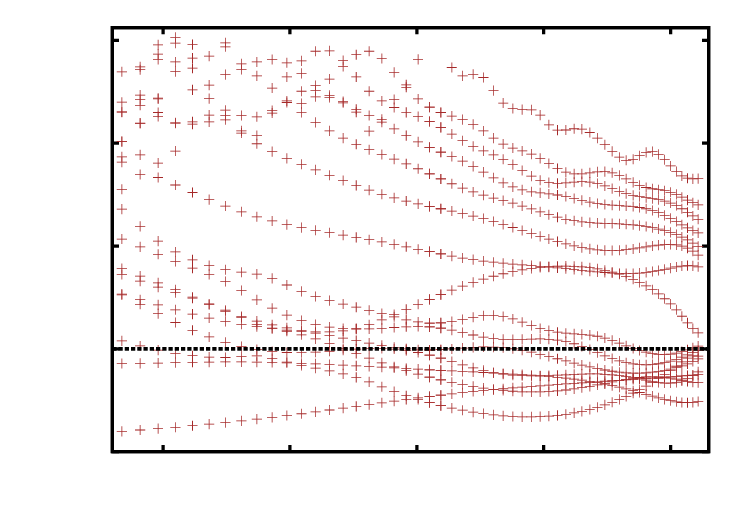} } \\
 \end{array}$
 \caption{Result of linear stability analysis in models of \textit{AF And} - Real (left) and 
 imaginary (right) parts of the eigenfrequencies are displayed as
a function of surface temperature . These eigenfrequencies are normalized with global 
free fall timescale. Negative imaginary parts indicate
unstable modes. Blue dots in the real part of eigenfrequency correspond to unstable modes. 
Several modes become unstable in models having
lower surface temperature.}
 \normalsize
 \label{AF}
 \end{figure*}

\begin{figure*}
\centering $
\large
\begin{array}{cc}

   \scalebox{0.65}{ \input{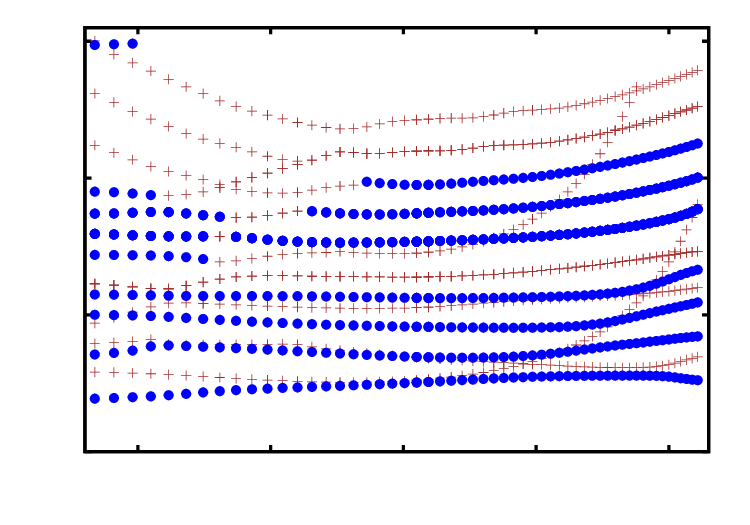} } 
   \scalebox{0.65}{ \input{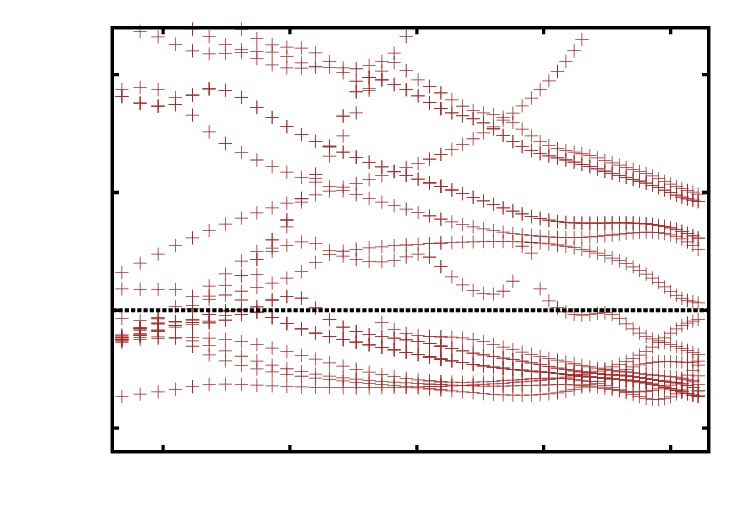} } \\
 \end{array}$
 \caption{Same as Fig. \ref{AF} but for models of  R\,127.}
 \normalsize
 \label{R127}
 \end{figure*}

Results of the performed linear stability analysis for models of \textit{AF And} and R\,127 are given in the form of 
`Modal diagram' in Fig. \ref{AF} and Fig. \ref{R127}, respectively. In the modal diagram, eigenfrequencies are given 
as a function of stellar parameter such as mass or surface temperature. Reader is 
referred to \citet{saio_1998} and \citet{gautschy_1990b} for 
detailed discussion on modal diagram. For models of \textit{AF And}, real and imaginary parts of eigenfrequencies 
as a function of surface temperature are given in Fig. \ref{AF}. Blue dots in the real part (Fig. \ref{AF}-left) and negative 
imaginary parts (Fig. \ref{AF}-right) are representing the unstable or excited modes. We note that all the considered models
of \textit{AF And} having surface temperature in the range of 32000 K to 9000 K are unstable. Two of the low order modes 
are excited in all the models. Number of unstable modes in models having temperature close to 32000 K are only two 
while models having temperature 9000 K have at least eight unstable modes. Models having lower surface 
temperature tend to have more unstable modes. Several mode interactions in terms of avoided crossings as 
well as instability bands are found in this modal diagram .


Similar to the case of \textit{AF And}, eigenfrequencies as a function of surface temperature are given in Fig.\ref{R127} for R\,127.   
In these models we also find several unstable modes. Analogous to \textit{AF And}, all the models of R\,127 in 
the considered surface temperature range are unstable. Four modes are unstable in all models. Mode interaction 
phenomena is also present in the modal diagram of R\,127 (Fig. \ref{R127}).

%\subsubsubsection{Subsubsubsection -- Level 4 title (Times New Roman 12 pts)}

%References are listed alphabetically by the first author's last name, with in-text citation in (Author, Year) style. Consistency throughout the reference list is required.

%Here some actual \LaTeX-style citation examples.
%For the sake of consistency, all authors are required to use the \texttt{natbib} package with the adequate bibliography style specified by the \texttt{bibliographystyle} command. Please make sure that the abbreviations for journal names are compliant with the usual conventions (see \texttt{JournalAbbreviations.txt}).

%Some citations have been included in the manuscript text with \texttt{natbib}'s citation command \verb|\citep{paper1,paper2}| and appear in the text as \citep{paper1,paper2}.
%This class is completed by the following {Bib\TeX} style file: \verb|bullsrsl-en.bst|. Another citation example is \citet{paper3}.

\section{Discussion and Conclusion} \label{d_and_c}
Unsteady mass-loss, irregular variability and surface eruptions have been observed in many LBVs. It has 
been pointed out in several studies that the models of LBVs are subject to dynamical instabilities which 
may be responsible for the observed irregular variabilities and surface eruptions
\citep[see e.g.][]{glatzel_1993c}. In the present analysis, 
we have considered models of two LBVs \textit{AF And} and R\,127. Linear stability analyses have been performed in 
model-sequence
having temperature in the range of 32000 K to 9000 K. Models having surface temperature close to 32000 K are 
representing the LBVs in quiescent phase while models with surface temperature close to 9000 K are 
associated with outburst phase. In the case of \textit{AF And}, we find more unstable modes in the model having surface 
temperature of 9000 K. Therefore it is an indication that as the star is approaching towards lower 
temperature (below 15000 K), more instabilities are appearing which may eventually lead to irregular 
variabilities and surface eruptions.  
However the final fate of instabilities can not be determined by 
linear stability analysis. Nonlinear numerical simulations is required to find out the final fate of 
the instabilities found in the considered models. For R\,127, strength of instabilities for some of the 
unstable models are increasing as the surface temperature of models are approaching towards 9000 K. Therefore it 
will be worth to explore the consequence of these instabilities. From the mode interactions in the both modal diagrams (Fig. \ref{AF} and Fig. \ref{R127}), we infer that the
several unstable modes are strange modes as found in earlier studies of massive stars \citep[see e.g.,][]{saio_1998, glatzel_1993,glatzel_1993c}.

Present work is a preliminary study to explore the role of instabilities in models of two LBVs.
More extensive linear stability analysis considering wide range of stellar mass and chemical composition 
followed by nonlinear numerical simulation for these two LBVs is the objective of ongoing work and the outcomes 
will be presented in the near future. Observational studies of LBVs combined with modelling using facilities of 
Belgo-Indian Network for Astronomy and Astrophysics (BINA) can enhance our present 
understanding about LBVs.

\begin{acknowledgments}
Authors acknowledge financial support from Science and Engineering Research Board (SERB), India through 
Core Research Grant (CRG/2021/007772). We thank the referee (Prof. Marc-Antoine Dupret) for his valuable suggestions to improve this paper.

\end{acknowledgments}

\begin{furtherinformation}

\begin{orcids}

\orcid{0000-0001-8262-2513}{Abhay Pratap}{Yadav}
%\orcid{1111-2222-3333-4444}{Leonie}{van Leon}
%\orcid{2222-3333-4444-5555}{Lotta}{Lothardis}

%{\sl This section is optional.
%You may list here the ORCIDs of those authors who would like to share them, 
%one per line, with the \verb|\orcid{|\texttt{\emph{ORCID}}\verb|}{|\texttt{\emph{First name}}\verb|}{|\texttt{\emph{Last name}}\verb|}| command.
%This command typesets the information, and makes the ORCIDs themselves active 
%links to the corresponding records on \href{https://orcid.org}{orcid.org}.

%Unlike in this sample, no other text should actually be included here and this section should reduce to a bare list.
%The \verb|\orcid| command controls line feeds by itself; please do not insert any \verb|\\| or \verb|\newline| before or after them.}
\end{orcids}

\begin{authorcontributions}
Authors have contributed equally.  
%This section is mandatory when there is more than one author.
%The contributions of each author (identified by their initials) must be declared.
%We recommend to follow the \href{http://credit.niso.org}{CRediT} taxonomy (Contributor Roles Taxonomy).
\end{authorcontributions}

\begin{conflictsofinterest}
%This section is \emph{mandatory}.
%Authors must declare any personal or professional circumstances that may be perceived as influencing the research reported in the paper.
%If there is no conflict of interest, please state that ``The authors declare no conflict of interest.''
The authors declare no conflict of interest.
\end{conflictsofinterest}

\end{furtherinformation}

\bibliographystyle{bullsrsl-en}

\bibliography{extra}

%\bibliography{first}
\end{document}